\documentclass{PoS}
\newcommand{\Fermi}{\textit{Fermi}}

\title{Morphological and spectral measurements of 2HWC J1928+177 with HAWC and H.E.S.S.}

\ShortTitle{Morphology and spectrum of 2HWC J1928+177}

\author{\speaker{R. L\'opez-Coto}$^a$, V. Marandon$^a$, F. Brun$^b$ for the HAWC$^c$ and HESS collaborations\\
        $^a$Max Planck Institut f\"ur Kernphysik, Heidelberg, Germany \\
        $^b$ Universit\'e Bordeaux, CNRS/IN2P3, Centre d'\'Etudes Nucl\'eaires de Bordeaux Gradignan, 33175 Gradignan, France \\
        $^c$For a complete author list, see http://www.hawc-observatory.org/collaboration/icrc2017.php
        E-mail: \email{rlopez@mpi-hd.mpg.de}}

\abstract{2HWC J1928+177 is a source recently discovered at TeV energies in the second HAWC catalog. It is coincident with the Fermi unidentified source 3FGL J1928+1739 and the pulsar PSR J1928+1746, which is 83 kyr old and located at a distance of 5 kpc with an Edot=1.6 x 10$^{36}$ erg/s. 2HWC J1928+177 is not detected by any of the IACTs currently in operation, which puts strong constrains on the morphology and spectrum of the source. There is no sign of shell-like structure in the region at other wavelengths, which in addition to the presence of a pulsar at the center, points to a Pulsar Wind Nebula origin of the TeV emission. We present a dedicated morphological and spectral analysis of the region using HAWC and H.E.S.S. data to unveil the nature of the source and study its properties.}

\FullConference{35th International Cosmic Ray Conference, ICRC2017 -\\
		10-20 July, 2017\\
		Bexco, Busan, Korea}

\begin{document}

\section{Introduction}
PSR J1928+1746 is a radio pulsar discovered in the ALFA survey \cite{Cordes2006}. It has a period of $P = 68.7$ ms and period derivative $\dot{P}=1.3209\times10^{-14}$ s s$^{-1}$, its characteristic age is $\sim$82 kyr and its spin-down power is $\dot{E}=1.6 \times 10^{36}$ erg s$^{-1}$.
It is located at a distance of 5.8 kpc derived from its Dispersion Measure and presents a flat radio spectrum \cite{Nice2013}. 
Chandra observed the pulsar as a part of their CHAPS survey \cite{chaps2012}. They established an upper limit on the pulsed emission of $5.8\times10^{-15}$ erg s$^{-1}$cm$^{-2}$ in the 0.5-8 keV range. 
Nothing seen in the Swift-XRT survey of \emph{Fermi} unassociated sources either \footnote{http://www.swift.psu.edu/unassociated/source.php?source=3FGLJ1928.9+1739}.

EGRET discovered  3EG J1928+1733, an unidentified source shown in \cite{Torres2001}. This source is coincident with the \Fermi\ unidentified source 3FGL J1928.9+1739 \cite{3fgl}, which is a steady source in the energy range between 100 MeV and 300 GeV. According to \cite{SazParkinson2016}, 3FGL J1928.9+1739 is classified as a likely pulsar. There is no source detected by the \Fermi\ satellite at higher energies \cite{3fhl}. 
The region was observed by VERITAS \cite{Acciari2010} as part of the obsevations of SNR G054.1+0.3. They established upper limits on the integral flux above 1 TeV of a steady source centered at PSR~J1928+1746 at a level $F < 2.6 \times 10^{-13}$ cm$^{-2}$ s$^{-1}$.

HAWC discovered 2HWC J1928+177 in the second HAWC catalog \cite{2hwc2017}. The source coincident with PSR J1928+1746 and 3FGL J1928.9+1739, but it is located in a very crowded region with the detection of 2HWC J1930+188 (SNR G054.1+0.3) and additional emission surrounding these two objects.
HESS observed the region as part of the HESS Galactic Plane Survey (HGPS) \cite{hgps}. HESS detected SNR G054.1+0.3, although there is no report of Very High Energy (VHE) $\gamma$-ray emission coincident with 2HWC J1928+177. 

In this contribution we will use HAWC measurement of 2HWC J1928+177 and the upper limits placed by HESS on the same region to unveil the nature of the source discovered by HAWC.

\section{Instruments and Data}
The HAWC Gamma-Ray Observatory is located at Sierra Negra, Mexico at 4100 m a.s.l., and is sensitive to gamma rays and cosmic rays in the energy range from 100 GeV to 100 TeV \cite{HAWC_Performance}.
It is composed by 300 optically isolated tanks covering an area of 22, 000 m$^2$. Each one of these Water Cherenkov  Detectors  (WCD)  consists  of  a  metallic  cylinder  of  7.3 m  diameter  and  4.5 m  height containing 180, 000 liters of water. They are equipped with one 10" PMT at the center and three 8" PMTs surrounding the central one. The array has a 2 sr field of view with $>$95\% uptime. HAWC started operation in its full configuration in March 2015. HAWC's one-year sensitivity is better than the 50-hour sensitivity of the current generation of imaging atmospheric Cherenkov telescopes (IACTs) for energies larger than 10 TeV. The results presented on this contribution are for the time range between 26th November 2014 and 18th February 2017, containing a total of 760.3 days of livetime data. The data are binned according to the fraction of the detector hit ($fHit$) and all the gamma/hadron separation cuts, reconstruction and significance calculations are performed using these bins \cite{HAWC_Performance}.

H.E.S.S. is an array of five IACTs located at an altitude of 1800 m above sea level in the Khomas highlands of Namibia. During its first phase, also known as HESS-I, the instrument was composed of four 12-m diameter mirror Cherenkov telescopes \cite{hess}. During its second phase or HESS-II, an additional 28-m diameter mirror Cherenkov telescope was built at the center of the array. The angular resolution achieved by the system is $<0.1^\circ$ and the energy resolution $\sim$15 \%. The HGPS uses data of the 4-telescope HESS system taken from 2004 to 2013 analyzed using a pipeline analysis. HESS took $\sim$30 hours of data in the 2HWC J1928+177 region.

\section{Results}
\label{sec:results}
The Test Statistic (TS) of the HAWC detection is TS=102.4, corresponding to a significance of $\sim10\sigma$ pre-trial significance assuming a point-like source and using a single power-law of the form:
\begin{equation}
\label{powerlaw}
 \frac{dN}{dE} = f_0 \left( \frac{E}{E_0} \right)^{-\Gamma}
\end{equation}

\begin{figure}
\begin{center}
\includegraphics[width=0.7\textwidth]{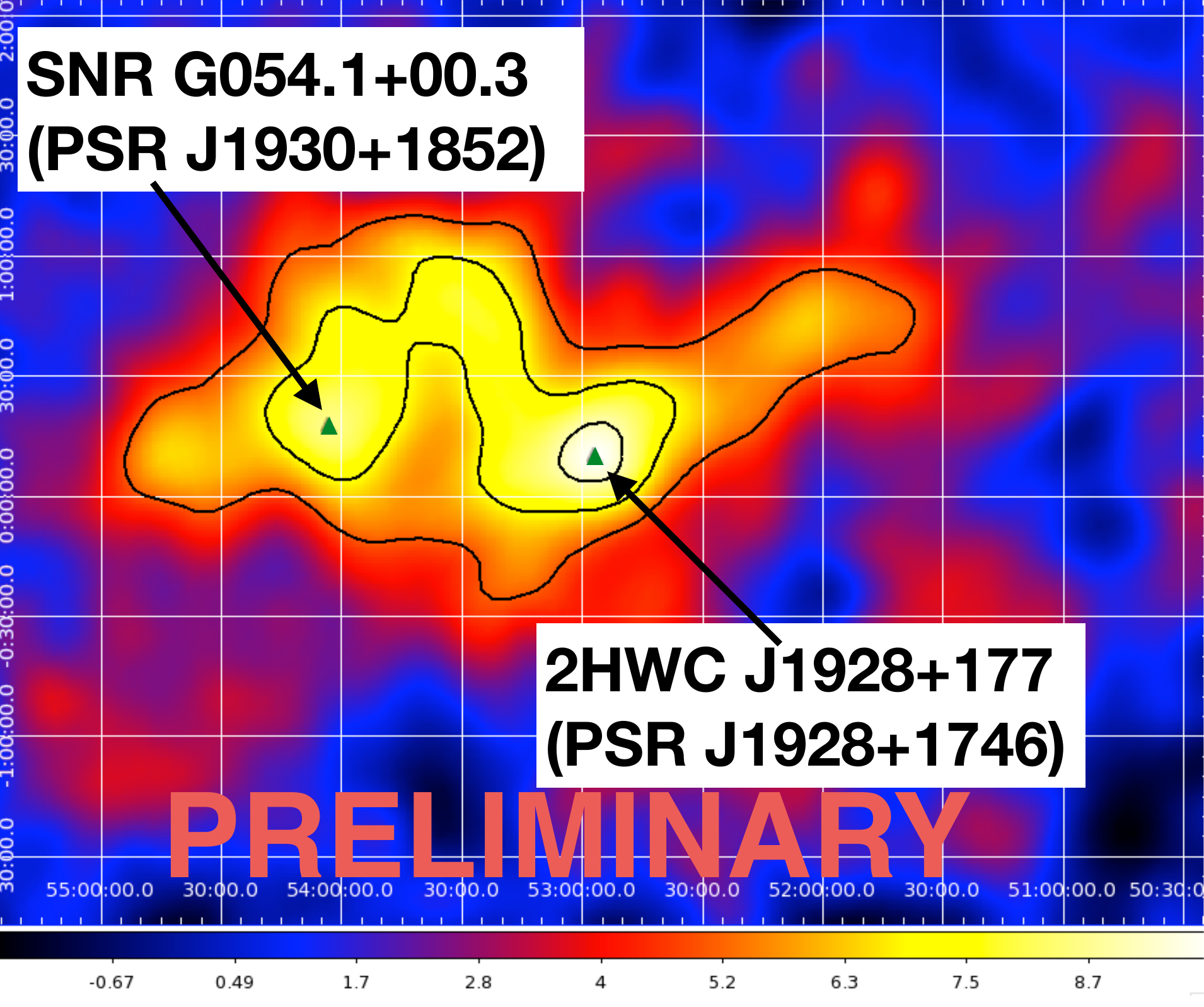}
\caption{Skymaps of the 2HWC J1928+177 region for all $fHit$ bins. Color scale corresponds to sqrt(TS). Black contours at sqrt(TS)=5, 7, 9. Green triangles correspond to the position of PSR J1930+1852 (left) and PSR J1928+1746 (right).}
\label{fig:skymap_all_bins}
\end{center}
\end{figure}

The fit parameters are $E_0$=7 TeV, $f_0=(1.07\pm0.12)\times10^{-14}$ TeV$^{-1}$ cm$^{-2}$ s$^{-1}$ and $\Gamma=2.60\pm0.09$. Following a method similar to the one used to determine the energy range in the 2HWC catalog, the energy range for which this fit is valid ranges from $\sim$1 TeV - 86 TeV. The source was detected with a significance $>3\sigma$/bin only for $fHit > 4$, which shifts the energy range where we confidently detect it to energies of a few TeV.  The skymap of the region using all the data available is shown on Figure \ref{fig:skymap_all_bins}.

\begin{figure}
\begin{center}
\includegraphics[width=0.7\textwidth]{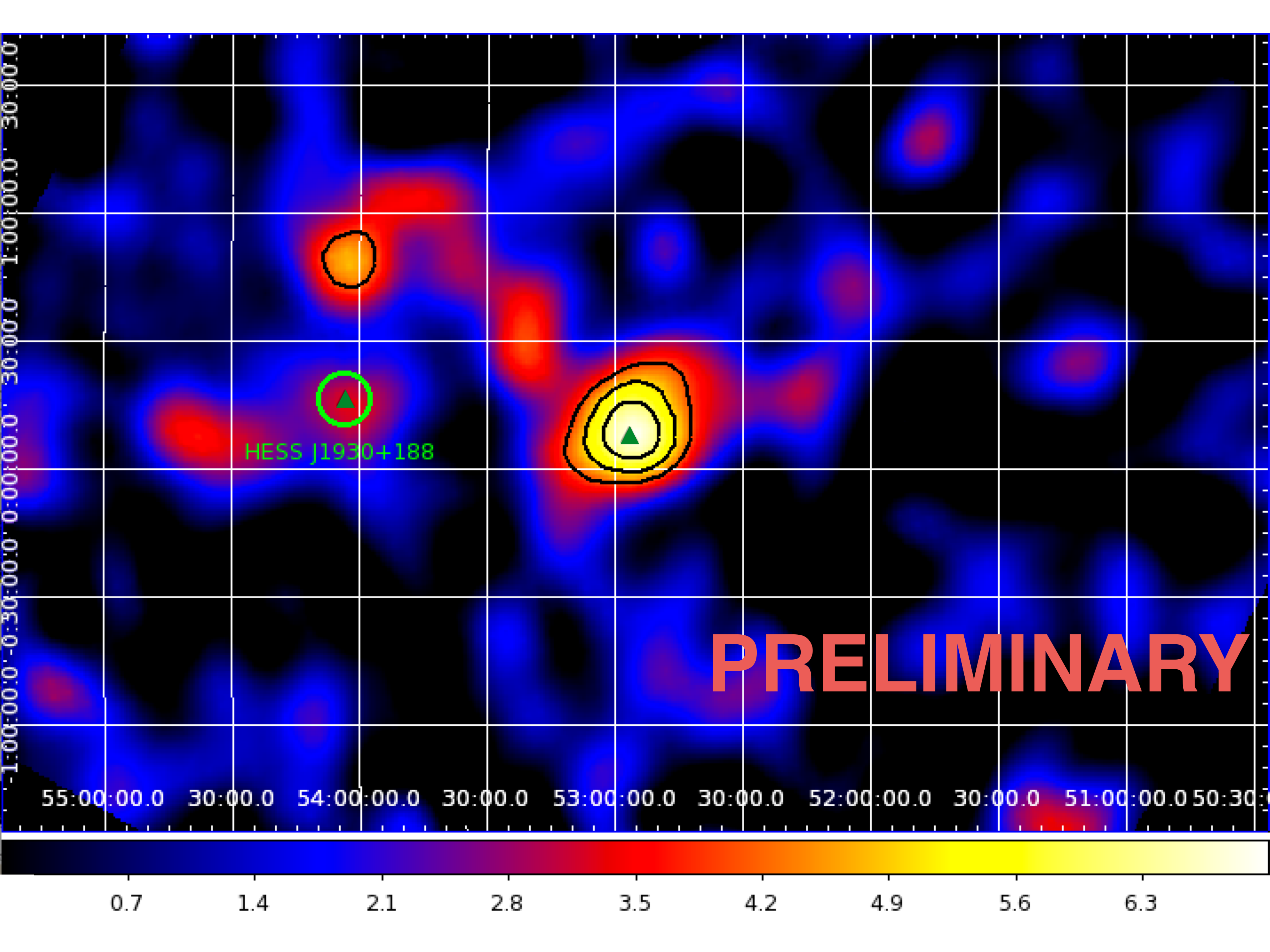}
\caption{Skymap of the 2HWC J1928+177 region for the $fHit$ bins 8 and 9. Color scale corresponds to sqrt(TS). Black contours at sqrt(TS)=4, 5, 6. Green triangles correspond to the position of PSR J1930+1852 (left) and PSR J1928+1746 (right).}
\label{fig:skymap_bins89}
\end{center}
\end{figure}

In order to get an improved angular resolution, we performed the analysis of the highest $fHit$ bins, where the angular resolution is $\sim0.2^\circ$. The skymap of the region is shown on Figure \ref{fig:skymap_bins89}. The energy range corresponding to these bins is $\geq 10$ TeV. The favored size for the source in this case is the point-like assumption of $0.2^\circ$.

\begin{figure}
\begin{center}
\includegraphics[width=0.8\textwidth]{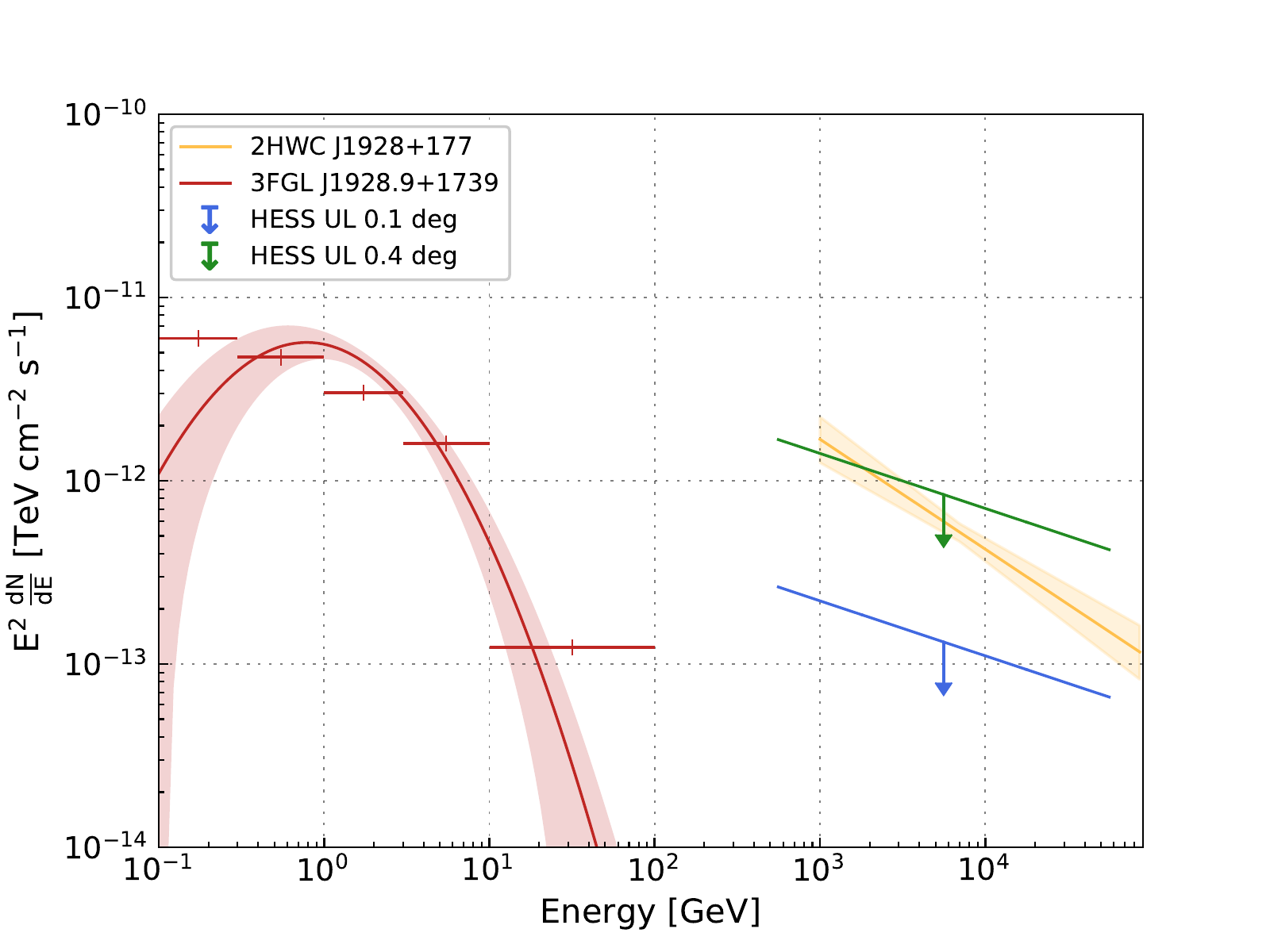}
\caption{Spectral energy distribution of 2HWC J1928+177 and 3FGL J1928.9+1739.}
\label{fig:sed_1928}
\end{center}
\end{figure}

In Figure \ref{fig:sed_1928}, we compare the spectrum measured by HAWC with the upper limits for a 95\% confidence level, calculated above a safe threshold ($\sim$ 500 GeV) for a HESS analysis using 0.1$^\circ$ and 0.4$^\circ$ integration radii. We also include the spectrum of 3FGL J1928.9+1739 as measured by Fermi. 

\section{Discussion}
Using the results presented in Section \ref{sec:results}, we discuss the origin of the VHE $\gamma$-ray emission of 2HWC J1928+177. Since there is no sign of unpulsed radio or X-ray source, and the 3FGL source coincident with the HAWC emission does not spectrally match the HAWC measured spectrum, we will use the VHE $\gamma$-ray measured spectrum and upper limits for the discussion. 

\subsection{Parent particle population}
Having a pulsar at the center and not showing any variability, the source is likely to belong to the Pulsar Wind Nebula (PWN) or Supernova Remnant (SNR) class. In the case of PWNe, the VHE $\gamma$-ray emission is originated by inverse Compton emission from electrons upscattering ambient photon fields \cite{aharonian}. The $\gamma$-ray emission produced in SNRs is believed to be produced by $\pi^0$ decay produced as the result of proton collisions \cite{pi0}. To calculate the parent particle population spectrum producing the observed $\gamma$-ray emission, we use the \texttt{naima} package \cite{naima}.

We first assumed that the emission is produced by electrons upscattering Cosmic Microwave Background (CMB) and Far Infrarred (FIR) photons with a 20 K temperature. The energy density of the target photon fields is 0.25 eV/cm$^3$ and 0.3 eV/cm$^3$ respectively. Since the HAWC measured spectrum does not show any cut-off, the best fit function for the electrons producing the VHE $\gamma$-ray emission is a single power-law of the same form as in equation \ref{powerlaw} with $E_0$=1 TeV, $f_0$=2.4$^{+0.7}_{-0.1}\times$10$^{47}$ erg$^{-1}$ and $\Gamma=3.25^{+0.09}_{-0.10}$. The total energy in electrons above 1 TeV that fits the VHE $\gamma$-ray data is $W_e$=4.9$^{+1.6}_{-1.2}\times$10$^{47}$ erg. The total energy released by the pulsar assuming an initial spin-down timescale of $\tau=10^4$ yr and a braking index of $n$=3 is $\sim10^{51}$ erg, meaning that only a small fraction of the energy injected by the pulsar needs to be invested into the acceleration of electrons to produce the VHE $\gamma$-ray emission we detect. 

We also assumed that the emission is generated by $\pi^0$ decay produced by proton collisions. Since there is no observational evidence of high density gas in the region, the assumed density of the medium is $n$ = 1 cm$^{-3}$. The spectrum is best fit by a power-law function with $E_0$=1 TeV, $f_0$=$(3.7^{+0.7}_{-1.0})\times10^{49}$ erg$^{-1}$ and $\Gamma=2.55^{+0.05}_{-0.08}$. The total energy needed to accelerate protons above 1 TeV is $W_p$=1.72$^{+0.13}_{-0.3}\times$10$^{50}$ (n/1 cm$^{-3}$)erg. Taking into account that the canonical value for the energy released by a Supernova (SN) explosion is $\sim10^{51}$ erg, it means that a factor $>$10\% of the energy of the SN needs to be transformed into acceleration of protons above 1 TeV to match the observed $\gamma$-ray emission. This is challenging, specially taking into account that the density of the medium might be smaller than the value considered, as it is the case for most of the VHE $\gamma$-ray emitting SNRs \cite{hess_snr}. Notice that a spectrum originated by hadronic emission and showing no cut-off up to $E=86$ TeV would also imply a hard spectrum not seen in any SNR at these energies yet.
A higher flux due to an extension of the emission up to larger angular distances would strengthen the case that protons are unlikely to produce the measured VHE $\gamma$-ray emission.

\begin{figure}
\begin{center}
\includegraphics[width=0.48\textwidth]{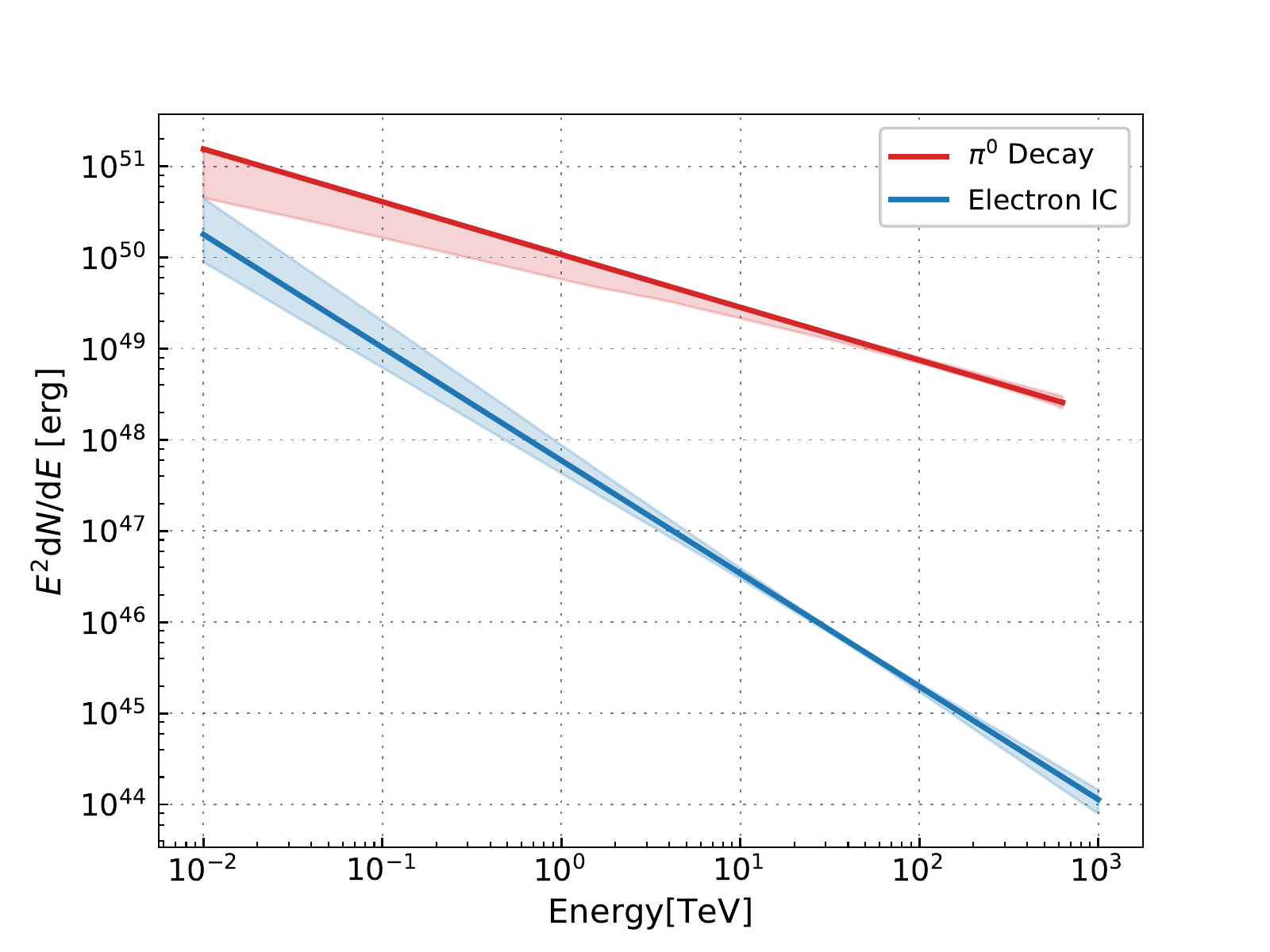}
\includegraphics[width=0.48\textwidth]{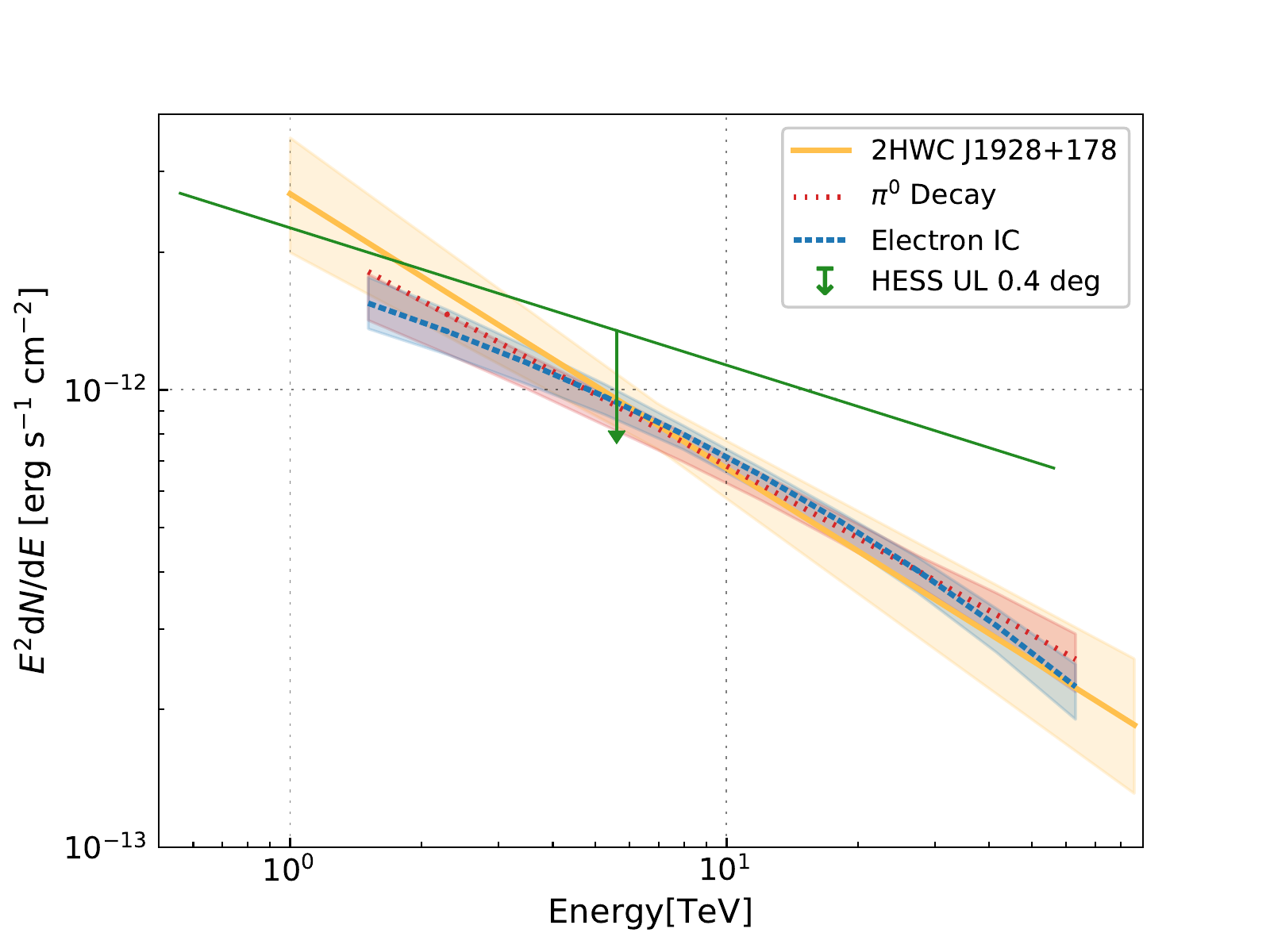}
\caption{\textbf{Left panel:} Particle population producing the observed VHE $\gamma$-ray emission. In red the proton spectral energy distribution, in blue the electron one. The shaded regions correspond to the 68\% uncertainties in the fit. \textbf{Right panel:} HESS upper limits, VHE $\gamma$-ray emission measured by HAWC, and produced by the electron and proton populations from the right panel.}
\label{fig:particle_gamma_spectra}
\end{center}
\end{figure}

In Figure \ref{fig:particle_gamma_spectra}, we show the spectral energy distribution of the particles producing the VHE $\gamma$-ray emission, compared to HAWC measurements and HESS upper limits.

\subsection{Morphology}
HESS upper limits are compatible with HAWC measurement for a size of the source of 0.4$^\circ$, which is the minimum size the source should cover at these energies to be compatible with the HESS upper limits. We note that since HESS upper limits were calculated above a safe threshold, it is not incompatible that the source is smaller at multi-TeV energies. Another possibility is that the spectral index of the source hardens below a few TeV and that is the reason for the non-significant detection in the lowest analysis bins. In this case, the source could still be smaller than 0.4$^\circ$ without being incompatible with HESS upper limits. The HAWC result points to a point-like (0.2$^\circ$) morphology at the highest energies ($>$ 10 TeV). This might be a hint of energy-dependent morphology, pointing to a PWN origin of the VHE $\gamma$-ray emission, but the errors on the highest energy morphology of the source are so large that they make it compatible with a largest size as well.

\subsection{Phenomenological model of PWNe}
Following the guidelines from the most recent paper on the PWN population based on their phenomenological properties\cite{hess_pwn} and assuming that the pulsar powering the nebula is PSR J1928+1746, we calculated what is the position of this source with respect to the fits of TeV efficiency, luminosity and photon index as a function of the spin-down power of the central pulsar and its characteristic age. We also compared what is the position of the source in the TeV extension and surface brightness fits (assuming a 0.5$^\circ$ size). We find that the assumption that the VHE $\gamma$-ray emission measured by HAWC has a PWN origin powered by the pulsar PSR J1928+1746 is in agreement with all the confidence intervals of these fits.

\section{Conclusion}
We studied the morphology and spectrum of the 2HWC J1928+177 to unveil the origin of the VHE $\gamma$-ray emission discovered by HAWC. HESS upper limits and HAWC spectrum and morphology point to a PWN origin of the VHE $\gamma$-ray emission. If confirmed by radio or X-ray measurements, this source would be the confirmed PWN powered by the oldest pulsar known. Sources of this age could be the connection between classical and confined PWNe and Geminga-like sources also known as electron halos, where the VHE $\gamma$-ray emission is not produced by confined particles, but by electrons diffusing into the interstellar medium.


\section*{Acknowledgements}
We	acknowledge	the	support	from:	the	US	National	Science	Foundation	(NSF);	the	
US	Department	of	Energy	Office	of	High-Energy	Physics;	the	Laboratory	Directed	
Research	and	Development	(LDRD)	program	of	Los	Alamos	National	Laboratory;	
Consejo	Nacional	de	Ciencia	y	Tecnolog\'{\i}a	(CONACyT),	M{\'e}xico	(grants	
271051,	232656,	260378,	179588,	239762,	254964,	271737,	258865,	243290,	
132197),	Laboratorio	Nacional	HAWC	de	rayos	gamma;	L'OREAL	Fellowship	for	
Women	in	Science	2014;	Red	HAWC,	M{\'e}xico;	DGAPA-UNAM	(grants	IG100317,	
IN111315,	IN111716-3,	IA102715,	109916,	IA102917);	VIEP-BUAP;	PIFI	2012,	
2013,	PROFOCIE	2014,	2015; the	University	of	Wisconsin	Alumni	Research	
Foundation;	the	Institute	of	Geophysics,	Planetary	Physics,	and	Signatures	at	Los	
Alamos	National	Laboratory;	Polish	Science	Centre	grant	DEC-2014/13/B/ST9/945;	
Coordinaci{\'o}n	de	la	Investigaci{\'o}n	Cient\'{\i}fica	de	la	Universidad	
Michoacana. Thanks to	Luciano	D\'{\i}az	and	Eduardo	Murrieta	for	technical	
support.

\end{document}